\documentclass[pra,aps,showpacs,twocolumn,floatfix]{revtex4}
\usepackage{graphicx}
\usepackage[ansinew]{inputenc}
\usepackage{array}
\usepackage{color}
\usepackage{amsmath}
\usepackage{amsxtra}
\usepackage{amstext}
\usepackage{amssymb}
\usepackage{latexsym}
\usepackage{dsfont}
\begin{document}
\newcommand{\beq}{\begin{eqnarray}}
\newcommand{\eeq}{\end{eqnarray}}
\newcommand{\doh}[1]{\dot{\hat {#1}}}
\newcommand{\hvc}[1]{\hat{\vec {#1}}}
\newcommand{\ddt}{\frac{d}{dt}}
\newcommand{\khz}{\ \mathrm {kHz}}
\title{The role of quantum fluctuations in the optomechanical properties of a Bose-Einstein condensate in a ring cavity}

\author{S. K. Steinke}
\author{P. Meystre}

\affiliation{B2 Institute, Department of Physics and College of Optical Sciences,\\
The University of Arizona, Tucson, AZ 85721 }

\begin{abstract}
We analyze a detailed model of a Bose-Einstein condensate trapped in a ring optical resonator and contrast its classical and quantum properties to those of a Fabry-P{\'e}rot geometry.  The inclusion of  two counter-propagating light fields and three matter field modes leads to important differences between the two situations. Specifically, we identify an experimentally realizable region where the system's behavior differs strongly from that of a BEC in a Fabry-P\'{e}rot cavity, and also where quantum corrections become significant. The classical dynamics are rich, and near bifurcation points in the mean-field classical system, the quantum fluctuations have a major impact on the system's dynamics.
\end{abstract}
\maketitle

\section{Introduction}
In recent years, there has been an explosion of interest in optomechanical systems in which at least one degree of freedom is cooled nearly to its quantum ground state~\cite{kippenberg2008, marquardt2009, oconnell2010, teufel2011}. In the top-down approach, the mechanical element (often one end-mirror of a Fabry-P\'{e}rot cavity that is allowed to oscillate) is initially in thermal equilibrium with its surroundings and then is cooled via radiation pressure. On the other hand, in the bottom-up approach, the mechanical portion of the system typically consists of ultracold atoms trapped inside a high-$Q$ optical resonator. The ultracold atomic system can be a thermal sample \cite{murch2008}, a quantum-degenerate Bose-Einstein condensate (BEC) \cite{brennecke2008, esslinger2009}, or even a quantum-degenerate gas of fermions~\cite{rina2010}. In the bottom-up situation the mechanical oscillator(s) are comprised of \textit{collective} momentum modes of the trapped gas, excited via photon recoil \cite{nagy2009,liu2009,keye2009,dan2009,aranya2009}. The dynamics of the collective interaction between photons and ultracold atoms have been studied in detail both theoretically and experimentally in the context of superradiance in BECs~\cite{Inouye1999,Schneble2003,Schneble2004,Yoshikawa2004,Moore1999, Piovella2001,Robb2005,Zobay2005,Uys2007} as well as collective atomic recoil lasing (CARL)~ \cite{Bonifacio1994,Bonifacio21994,Bonifacio1995,Bonifacio1997,Lippi1996,Hemmer1996,Kruse2003,Cube2004}. In both these situations, the interplay between light fields and atomic motion leads to feedback effects. The bottom-up approach to optomechanics exploits this interplay as well, with the added feature that the motion of the atoms is analogous to that of a mechanical element driven by radiation pressure.

In the case of a high-$Q$ Fabry-P{\'e}rot cavity the intracavity standing-wave field couples the macroscopically occupied zero-momentum component of the BEC to a symmetric superposition of the states with center-of-mass momentum  $\pm 2\hbar k$ via virtual electric dipole transitions~\cite{brennecke2008,esslinger2009}. As discussed in a previous paper \cite{chen2010}, there are situations where a ring cavity can lead to atomic dynamics different from the standing-wave situation. This is because in contrast to a standing wave, running waves permit one in principle to extract ``which way'' information about the matter-wave diffraction process. As a first step toward discussing that question, the earlier work considered the difference between {\em classical} standing wave and counterpropagating light fields, that is, the difference in optomechanical properties of condensates trapped in, say, a Fabry-P{\'e}rot and a ring cavity. One main consequence of the presence of two counterpropagating running waves was that in addition to a symmetric ``cosine'' momentum side mode, it becomes possible to excite an out-of-phase ``sine'' mode as well. In the optomechanics analogy, this indicates that two coupled ``condensate mirrors'' of equal oscillation frequencies but in general different masses are driven by the intracavity field. We showed that this can result in complex multistable behaviors, including the appearance of isolated branches of solutions for appropriate choice of parameters. 

The present paper builds on these results and includes two new features. At a classical (operators replaced by $c$-numbers) level, the evolution of the zero-momentum mode of the condensate is now also included. Furthermore, this work also discusses the role of small quantum fluctuations in the system, particularly on the occupancy of the sine and cosine side modes. As previously discussed, together with the original condensate they form an effective V-system, with the upper levels -- the sine and cosine modes -- driven by a two-photon process involving both counterpropagating light fields. At the classical level one or the other of these modes can become a dark state, but quantum fluctuations will normally prevent these modes from becoming perfectly dark~\cite{shore1990}. It follows that measuring correlation functions of the optical field provides a direct means to probe the quantum properties of the matter-wave side modes. These and other aspects of the role of quantum fluctuations are examined in the following sections, which consider the situation where these fluctuations are feeble and their effect can be treated in the framework of a linearized theory.

This paper is organized as follows: Section II introduces our model of a quantum-degenerate atomic system interacting with two quantized counter-propagating field modes in a high-$Q$ ring resonator, and casts it in a form that emphasizes the optomechanical nature of the problem. Section III derives the resulting Heisenberg-Langevin equations of motion for the system. It solves them first in steady state for the case of classical fields, recovering in a slightly different form some key results of Ref. \cite{chen2010}. Section IV turns to the study of the role of quantum fluctuations on the system dynamics. It starts by deriving the linearized equations of motion that govern these fluctuations, and then briefly outlines a general treatment of quantum correlations applicable in cases where many separate noise sources are present. The remainder of the section exploits these results to analyze the different behaviors produced by the interplay between classical mean field dynamics and quantum fluctuations for selected system parameters. Finally, Section V is a summary and conclusion.

\section{Model}
We consider a Bose-Einstein condensate of $N$ two-state atoms with transition frequency $\omega_a$ and mass $m$, assumed to be at zero temperature, confined along the path of two counterpropagating optical beams in a ring cavity of natural frequency $\omega_c$ and wave number $k_c=\omega_c/c$. These fields are driven by external lasers of intensity $\eta_i$ and frequency $\omega_p$. We assume that the atomic transition is far off-resonance from the field frequency, so that the upper electronic level can be eliminated adiabatically. Neglecting two-body collisions and in a frame rotating at the pump frequency $\omega_c$ the Hamiltonian for this system is then
\begin{equation}
\hat H = \hat H_{\rm opt}+\hat H_{\rm pump}+\hat H_{\rm BEC}+\hat H_{\rm int},
\end{equation}
where
\beq
\hat H_{\rm opt} &=& -\hbar\sum_{i=1}^2\Delta \hat a_i^\dagger \hat a_i,\nonumber\\
\hat H_{\rm pump} &=& i\hbar\sum_{i=1}^2\eta_i \hat a_i^\dagger -  \eta^*_i \hat a_i,\nonumber\\
\hat H_{\rm BEC} &=& \int dx\hat\psi^\dagger (x)\left(-\frac{\hbar^2}{2m}\frac{d^2}{dx^2}\right) \hat \psi(x),\nonumber\\
\hat H_{\rm int} &=& \hbar \Omega_0\int dx\hat\psi^\dagger (x)(\hat a_1^\dagger \hat a_1+\hat a_2^\dagger \hat a_2\nonumber\\
&+& \hat a_2^\dagger \hat a_1 e^{2ikx} + \hat a_1^\dagger \hat a_2 e^{-2ikx}) \hat \psi(x).
\eeq
Here $\Delta = \omega_p-\omega_c$ is the pump-cavity detuning,
\begin{equation}
\Omega_0 = g_0^2/(\omega_p-\omega_a)
\end{equation}
is the off-resonant vacuum Rabi frequency and $g_0$ is the usual (resonant) vacuum Rabi frequency.

The photon recoil associated with the virtual transitions between the lower and upper atomic electronic states results in the population of atomic center-of-mass states of momenta $2p\hbar k$, where $p$ is an integer. For feeble intracavity fields and large detunings it is sufficient to consider the first two momentum side modes, $p=\pm 1$. It is then convenient to decompose the atomic Schr{\"o}dinger field in terms of its momentum ground state and two nearest momentum side modes in terms of the parity (rather than momentum) eigenstates
\beq
\label{psiofx}
\hat\psi(x) &=& \sqrt{\frac{2}{L}}\left [ \frac{\hat c_0}{\sqrt 2} + \hat c_c \cos(2kx) + \hat c_s \sin(2kx)\right ],
\eeq
where ${\hat c_0}$, ${\hat c_c}$ and ${\hat c_s}$ are bosonic annihilation operators for the zero-momentum component and for the sine and cosine side modes of the quantum-degenerate atomic system, respectively.

To couch the problem in a more transparently optomechanical form, we further make the substitutions \footnote{Recasting the light field operators in terms of ``position'' and ``momentum'' operators has been done primarily for later analytical ease rather than as a straightforward analogy to other optomechanical problems.}
\beq
\label{c0}
\hat c_0 &=& \sqrt N + \frac{\hat X_0 + i \hat P_0}{\sqrt 2},\\
\hat a_{1,2} &=& \frac{\hat X_{1,2} + i \hat P_{1,2}}{\sqrt 2},\nonumber\\
\label{aici}
\hat c_{c,s} &=& \frac{\hat X_{c,s} + i \hat P_{c,s}}{\sqrt 2}.
\eeq
The first of these equations is indicative of the fact that we assume that the zero-momentum component of the atomic sample comprises a macroscopically populated component that we treat in mean-field theory via a classical amplitude $\sqrt N$, to which are superimposed quantum fluctuations resulting from the coupling to the sine and cosine side modes.

The approximate expansion (\ref{psiofx}) and the definitions (\ref{c0})-(\ref{aici}) result in the alternate form of the Hamiltonian
\begin{equation}
{\hat H}'= {\hat H}'_{\rm opt} + {\hat H}_{\rm pump}+{\hat H}'_{\rm BEC}+{\hat H}'_{\rm int}
\label{newH}
\end{equation}
where
\beq
\hat H^\prime_{\rm opt} &=& -\hbar\sum_i\frac{\tilde\Delta}{2}(\hat X_i^2+\hat P_i^2),\nonumber\\
\hat H_{\rm pump} &=& \hbar\sqrt 2\sum_i {\rm Re}(\eta)\hat P_i-{\rm Im}(\eta)\hat X_i,\nonumber\\
\hat H^\prime_{\rm BEC} &=& \hbar\frac{\omega_r}{2}(\hat X_c^2+\hat X_s^2+\hat P_c^2+\hat P_s^2),\nonumber\\
\hat H^\prime_{\rm int} &=&\hbar\frac{\Omega_0}{\sqrt 2}(\hat L_e\hat M_c +\hat L_o\hat M_s),
\eeq
and
\begin{eqnarray}
\hat L_e &=& \hat X_1 \hat X_2 + \hat P_1 \hat P_2,\nonumber\\
\hat L_o &=& \hat X_1 \hat P_2 - \hat X_2 \hat P_1,\nonumber\\
\hat M_I &=&\sqrt{2N}\hat X_I +\hat X_0\hat X_I+\hat P_0\hat P_I,
\end{eqnarray}
where $I=\{c,s\}$. In Eq.~(\ref{newH}) a constant term has been ignored and the energy shift of the atom-light interaction has been absorbed into the optical part of the Hamiltonian (hence the primed terms).

The operators $\hat L$ and $\hat M$ are quadratic light and matter operators, respectively, while the $e$ and $o$ subscripts in the light operators indicate their parity under interchange of the left- and right-moving light fields. Finally,
\begin{equation}
\tilde\Delta = \Delta - \Omega_0N
\end{equation}
is an effective detuning that accounts for the mean-field Stark shift of the condensate and
\begin{equation}
\omega_r = 2\hbar k^2/m
\end{equation}
is the recoil frequency associated with the virtual transition.

As already discussed in Ref.~\cite{chen2010} the presence of two counter-propagating fields in a ring resonator results in a situation that is significantly more complex than is the case for a high-$Q$ Fabry-P{\'e}rot cavity. In particular, the optomechanical properties of the condensate are now formally analogous to those of a system of two coupled moving mirrors. This can be seen by exploiting the fact that due to the large value of $N$, we can for now neglect the nonlinear terms in $\hat M_c$ and $\hat M_s$, so that
\begin{equation}
\hat M_i \simeq \sqrt{2N}\hat X_i,
\end{equation}
and hence
\begin{equation}
\hat H^\prime_{\rm int} \simeq{\Omega_0 \sqrt N}(\hat L_e\hat X_c +\hat L_o\hat X_s).
\end{equation}
Thus, rather than having a light-matter interaction proportional to the light field intensity times the position of an effective mirror we now have an interaction with two ``mirrors'' ~\cite{Bhattacharya2008} of equal mass and effective oscillation frequency, but each of which is driven differently due to interference effects between the two counterpropagating light fields.

\section{Mean-field steady state and stability}
The Heisenberg-Langevin equations of motion of the system are easily derived from the Hamiltonian~(\ref{newH}), complemented by appropriate quantum noise and damping terms. One finds readily
\beq
\doh X_i &=& -\kappa\hat X_i -\tilde\Delta \hat P_i + {\rm Re}(\eta)\sqrt 2 \nonumber\\
 &+& \frac{\Omega_0}{\sqrt 2}( \hat M_c\hat P_j +(-1)^i\hat M_s\hat X_j )+\hat\xi_{xi},\nonumber \\
\doh P_i &=& -\kappa\hat P_i +\tilde\Delta \hat X_i + {\rm Im}(\eta)\sqrt 2\nonumber\\
 &-& \frac{\Omega_0}{\sqrt 2}(\hat M_c\hat X_j +(-1)^i\hat M_s\hat P_j)+\hat\xi_{pi}, \nonumber\\
\doh X_0 &=& -\gamma\hat X_0 +\frac{\Omega_0}{\sqrt 2}( \hat L_e\hat P_c +\hat L_o\hat P_s)+\hat\xi_{x0},\nonumber\\
\doh P_0 &=& -\gamma\hat P_0 -\frac{\Omega_0}{\sqrt 2}(\hat L_e\hat X_c +\hat L_o\hat X_s )+\hat\xi_{p0},\nonumber
\eeq\beq    %%%% Split for aesthetics
\doh X_c &=& -\gamma\hat X_c + \omega_r \hat P_c + \frac{\Omega_0}{\sqrt 2}\hat L_e\hat P_0+\hat\xi_{xc},\nonumber\\
\doh P_c &=& -\gamma\hat P_c - \omega_r \hat X_c - \frac{\Omega_0}{\sqrt 2}\hat L_e(\sqrt{2N}+\hat X_0)+\hat\xi_{pc},\nonumber\\
\doh X_s &=& -\gamma\hat X_s + \omega_r \hat P_s + \frac{\Omega_0}{\sqrt 2}\hat L_o\hat P_0+\hat\xi_{xs},\nonumber\\
\label{EOM}\doh P_s &=& -\gamma\hat P_s - \omega_r \hat X_s - \frac{\Omega_0}{\sqrt 2}\hat L_o(\sqrt{2N}+\hat X_0)+\hat\xi_{ps},
\eeq
where $i=\{1,2\}$ and $j = 3-i$.

The noise sources are assumed uncorrelated for the different modes of both the matter and light fields. Because the damping originates in the $\hat a$ and $\hat c$ operators, it appears in both the $\hat X$ and $\hat P$ equations of motion. In the case of the light fields, the noise and damping originate from cavity loss, vacuum noise, and laser fluctuations, while for the matter fields the primary source of noise and damping is 3-body collisions with additional nearby non-condensed atoms. In addition, the customary factors of $\sqrt{2\kappa}$ and $\sqrt{2\gamma}$ multiplying the $\hat\xi$'s (see e.g.~\cite{walls2007}) have been absorbed into their definitions. This simplifies later results somewhat.

\subsection{Comparison to previous results}
\begin{figure}[t]
\includegraphics[width=0.45 \textwidth]{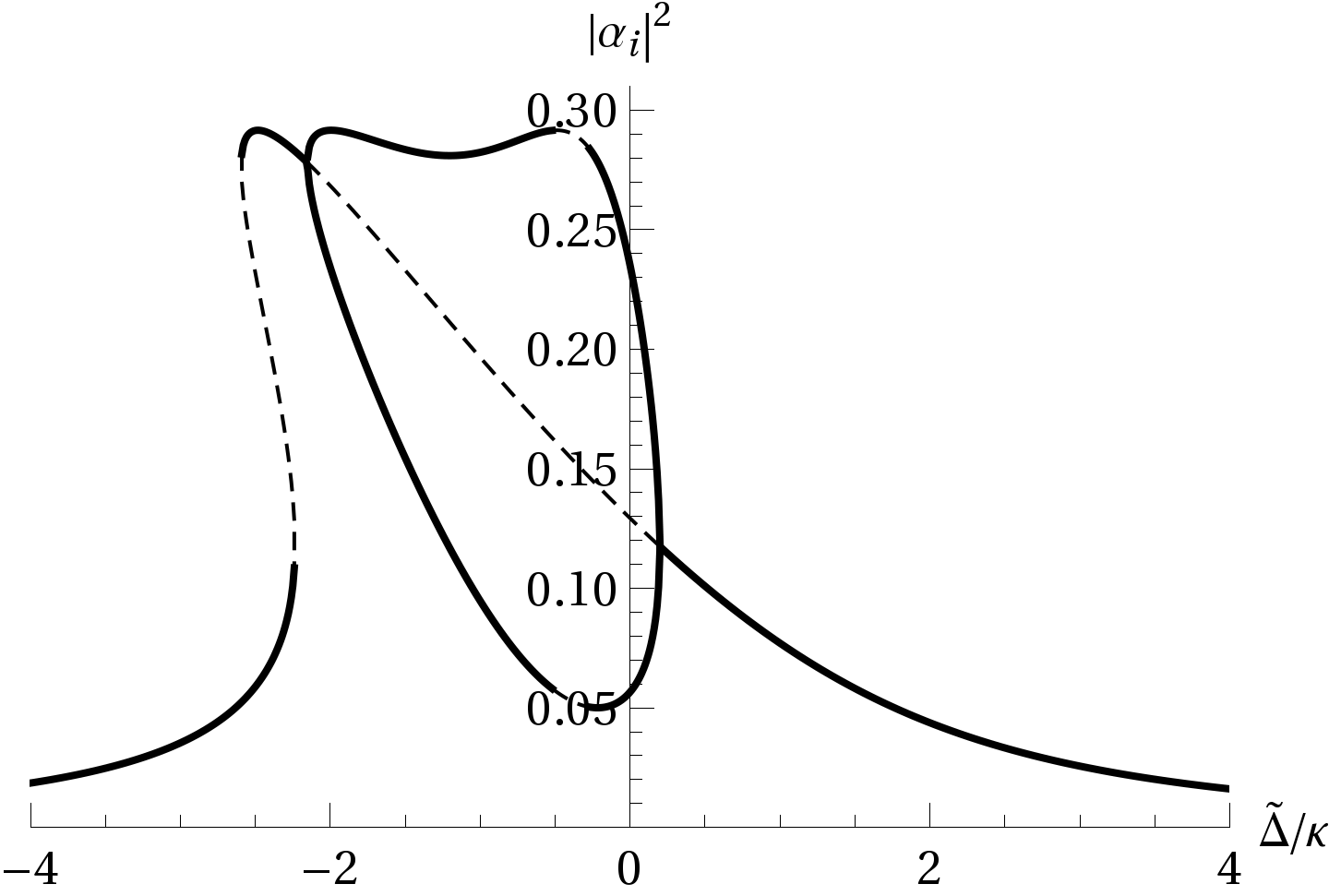}
\caption{\label{fig:bifurc}Mean intracavity photon number of both modes as a function of the effective detuning $\tilde\Delta$. Here $\kappa=2\pi\times1.3 {\rm~MHz}$, $\gamma = 2\pi\times1.3\khz$, $\Omega_0=2\pi\times3.1\khz$, $\omega_r = 2\pi\times15.2\khz$, $N=9000$, and $\eta_1=\eta_2=0.54\kappa$. Stable solutions are solid and unstable are dashed.}
\includegraphics[width=0.45 \textwidth]{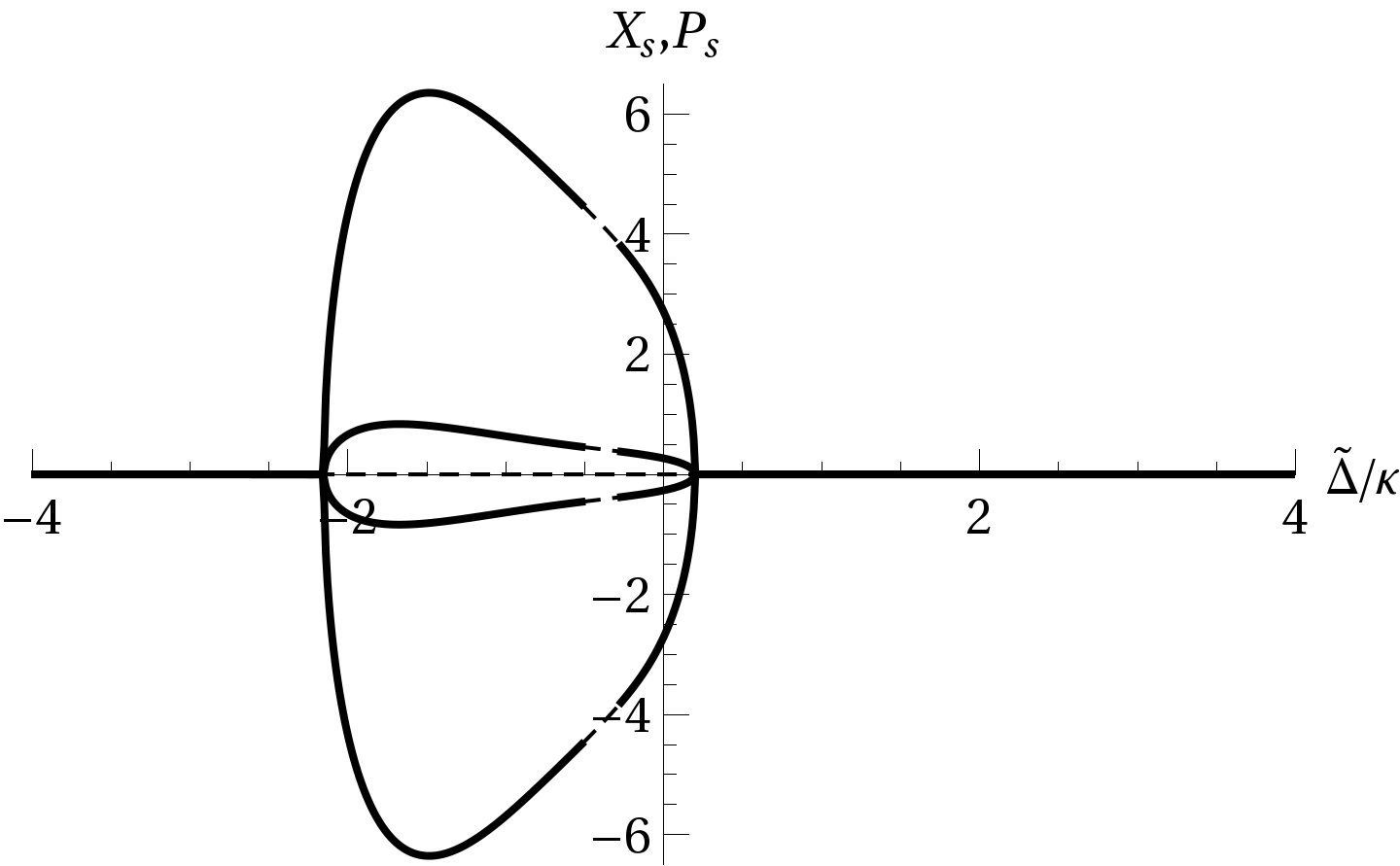}
\caption{\label{fig:xspsnoa}Mean sine mode ``position'' and ``momentum'' as a function of $\tilde\Delta$. Other parameters as above. The outer magnitude branches are the position solutions and the inner the momentum. The maximum sine mode occupancy $N_s$ reached is roughly 20.}
\end{figure}

Before undertaking an analysis of the role of quantum fluctuations, we investigate the effects of including the zero-momentum mode itself as a dynamical component in the {\em classical} steady-state of the system. Surprisingly, perhaps, we find that its inclusion can result in the appearance of new dynamical features, such as e.g. a Hopf bifurcation. To proceed we neglect all noise operators, treat the remaining fields classically in a standard mean-field approach, and simply look at the existence and stability of the fixed points of the classical version of Eqs.~(\ref{EOM}) as the various parameters are varied. The results of one such calculation are shown in Fig.~\ref{fig:bifurc}, revealing the dependence of the intracavity photon number $|\alpha_i|^2=\frac{1}{2}(X_i^2+P_i^2)$ on the detuning $\tilde\Delta$.  (We use the word ``photon'' rather loosely in the classical description of this section; more accurately it is a dimensionless measure of the mean light field intensity.)

When compared to Fig.~2 of Ref.~\cite{chen2010}, which uses identical parameters, we note that a quite similar bistable behavior is observed, with two degenerate stable branches the photon number can reach, as before. Which branch is reached is dependent on initial conditions and quantum fluctuations, and, when one field's intensity is given by the lower branch, the other's is given by the upper. There is one particularly interesting discrepancy when the evolution of the zero-momentum mode's occupancy is included, however. For a small range of negative detunings -- around $\tilde\Delta/\kappa \approx -0.5$ for the present example -- the bistable branches become unstable, due to a previously unseen Hopf bifurcation~\cite{tabor1989}. We attribute this bifurcation to the inclusion of the dynamics of the zero-momentum mode, an aspect that was neglected in the analysis of Ref.~\cite{chen2010}. Bistable periodic cycles are present, and as before, which is reached depends primarily on the initial state of the system. The amplitudes of oscillations observed in these stable cycles are quite small for the zero-momentum mode when compared to its mean value (which is of the order $\sqrt{N}$), but for the lightly occupied side modes, the amplitudes can be comparable to the mean values. This is a case where simply making the ansatz $\hat c_0 \rightarrow \sqrt{N}$ suppresses certain dynamical features. 

In order to do a meaningful linearized quantum treatment, we attempt in this paper to avoid parameter regions where complicated multistable behavior is evident. The results of the next section will assist us in this effort. Furthermore, a broader search of the parameter space has hinted that there are also experimentally realizable regimes in which the classical dynamics goes beyond multistability and becomes chaotic. We hope to return to this topic in later work.

\subsection{Classical dark state}
We now turn to a concrete example in which quantum and classical predictions will be shown to differ qualitatively. Specifically, we consider the case of symmetric pumping $\eta_1=\eta_2=\eta$ of the counterpropagating cavity modes. We have shown previously in Ref.~\cite{chen2010} and confirm in the following that within a classical (for the light field) and mean-field (for the atoms) theory the sine mode is a dark state, for all but relatively narrow parameter regions in which bistability and spontaneous symmetry breaking occurs (Fig.~\ref{fig:xspsnoa}). This can be seen easily by replacing all optical and matter-wave operators by classical expectation values, for instance
\begin{equation}
\hat X_0 \rightarrow \bar X_0,
\end{equation}
where the overbar indicates the mean. Similar definitions are used for the rest of the linear operators. When a quadratic operator is written with an overbar, we refer to the products of the classical means of its constituents ($\bar L_e=\bar X_1\bar X_2+\bar P_1\bar P_2$, $\bar N_c = (\bar X_c^2+\bar P_c^2)/2$, etc.).

The steady-state solution in this limit is easily obtained by setting all time derivatives equal to zero. Because the equations of motion are coupled cubic polynomials, there may be multiple real-valued solutions. Also, because there are 10 variables, we cannot necessarily expect to find analytic solutions. Nevertheless, we find that the dark sine mode solution always exists and has a closed form. For much, though not all, of parameter space, it turns out to be the unique, stable mean field solution.  A primary point of interest for the remainder of this work is to identify conditions such that quantum corrections lead to a finite occupancy of this mode, in particular in or near parameter regions where other solutions appear. Thus, we begin with the ansatz
\begin{equation}
\bar X_s = \bar P_s = 0.
\end{equation}
This allows the equations of motion for the matter field operators to be simplified, yielding constraints for the light field mean values:
\beq
\bar X_1 \bar X_2 + \bar P_1 \bar P_2 &=& 2|\alpha_D|^2\nonumber,\\
\bar X_1 \bar P_2 - \bar X_2 \bar P_1 &=& 0,
\eeq
where $|\alpha_D|^2$ is the mean number of intracavity photons in either of the light modes, assuming the dark sine mode ansatz holds. In other words, it is the projected number of photons in the cavity modes assuming that the sine mode is empty. For completeness and to make more direct contact with Ref.~\cite{chen2010}, we also solve for the other classical steady state values.
\beq
\bar X_0 &=& -\Omega^2(\gamma^2+\Omega^2)Z,\nonumber\\
\bar P_0 &=& \gamma^2 \Omega^2 Z,\nonumber\\
\label{xcclass}\bar X_c &=& -\gamma^2\omega_r\Omega Z,\nonumber\\
\label{pcclass}\bar P_c &=& -\gamma\Omega(\gamma^2+\Omega^2)Z,
\eeq
where
\begin{equation}
\label{defineOmega}\Omega=\Omega_0|\alpha_D|^2\sqrt 2 
\end{equation}
is the off-resonant Rabi frequency for the intracavity photon number $|\alpha_D|^2$ and
\begin{equation}
\label{defineZ}Z = \frac{\sqrt {2N}}{(\gamma^2+\Omega^2)^2+\gamma^2\omega_r^2}.
\end{equation}
We note that if we calculate $\bar M_c$ and $\bar M_s$, the quantities governing light-cosine mode and light-sine mode interaction strengths, we retrieve $ \bar X_c\sqrt{2N}$ and 0, which are exactly the results obtained when occupancy changes in the zero-momentum mode are neglected. This shows that allowing for the evolution of the zero-momentum mode as in Eq.~(\ref{c0}) versus fixing $\hat c_0 \rightarrow \sqrt{N}$ will only yield different behaviors for the light and side-mode fields in the spontaneously broken symmetry region of parameter space.

The remaining four equations are easily solved to give
\beq
\bar X_{1,2} &=& \frac{\kappa{\rm Re}(\eta)\sqrt 2 - (\tilde\Delta-\Omega_0\bar{X_c}\sqrt{N}){\rm Im}(\eta)\sqrt 2 }{\kappa^2+(\tilde\Delta-\Omega_0\bar{X_c}\sqrt{N})^2},\nonumber\\
\bar P_{1,2} &=& \frac{\kappa{\rm Im}(\eta)\sqrt 2 + (\tilde\Delta-\Omega_0\bar{X_c}\sqrt{N}){\rm Re}(\eta)\sqrt 2 }{\kappa^2+(\tilde\Delta-\Omega_0\bar{X_c}\sqrt{N})^2}.
\eeq
We may therefore eliminate $P_{1,2}$ without loss of generality (thereby requiring $\alpha_D=\bar a_{1,2}$ to be real) by appropriate selection of the real and imaginary parts (equivalently, the phase and strength) of the pumping $\eta$, namely
\beq
\label{etatoalpha}
{\rm Re}(\eta) &=& \kappa\alpha_D,\nonumber\\
{\rm Im}(\eta) &=& -(\tilde\Delta-\Omega_0\bar{X_c}\sqrt{N})\alpha_D.
\eeq
\begin{figure}[t]
\includegraphics[width=0.45 \textwidth]{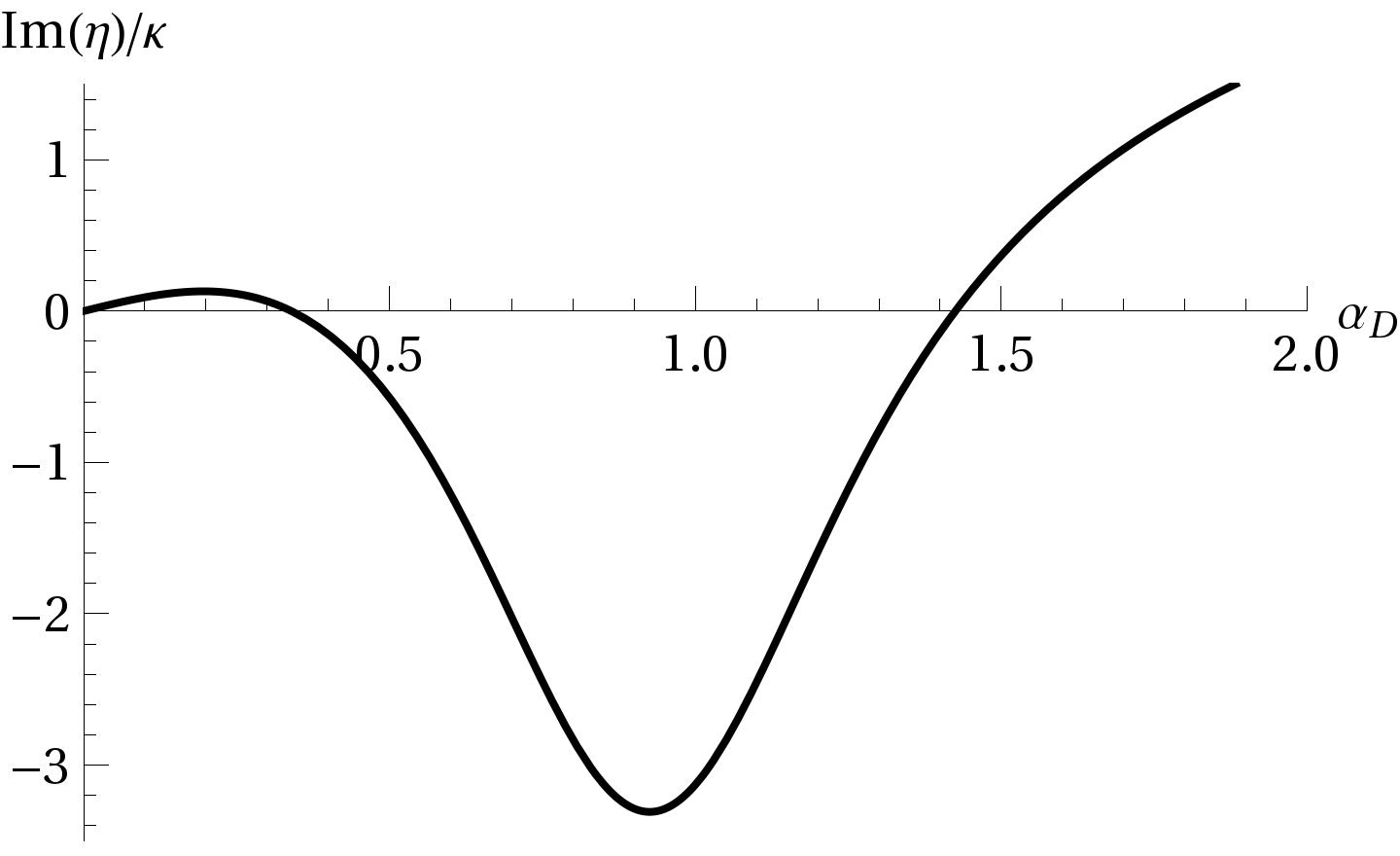}
\caption{\label{fig:etavalpha}Imaginary part of the pump intensity $\eta$ required to produce mean intracavity field $\alpha_D$. All parameters are as in Fig.~\ref{fig:bifurc}, except $\eta$ itself and $\tilde\Delta=-1.0\times\kappa$. Note that because ${\rm Re}(\eta)$ is linear in $\alpha_D$ (${\rm Re}(\eta)=\kappa\alpha_D$), $\eta$ itself is determined uniquely as a function of $\alpha_D$.}
\end{figure}
\begin{figure}[b]
\includegraphics[width=0.45 \textwidth]{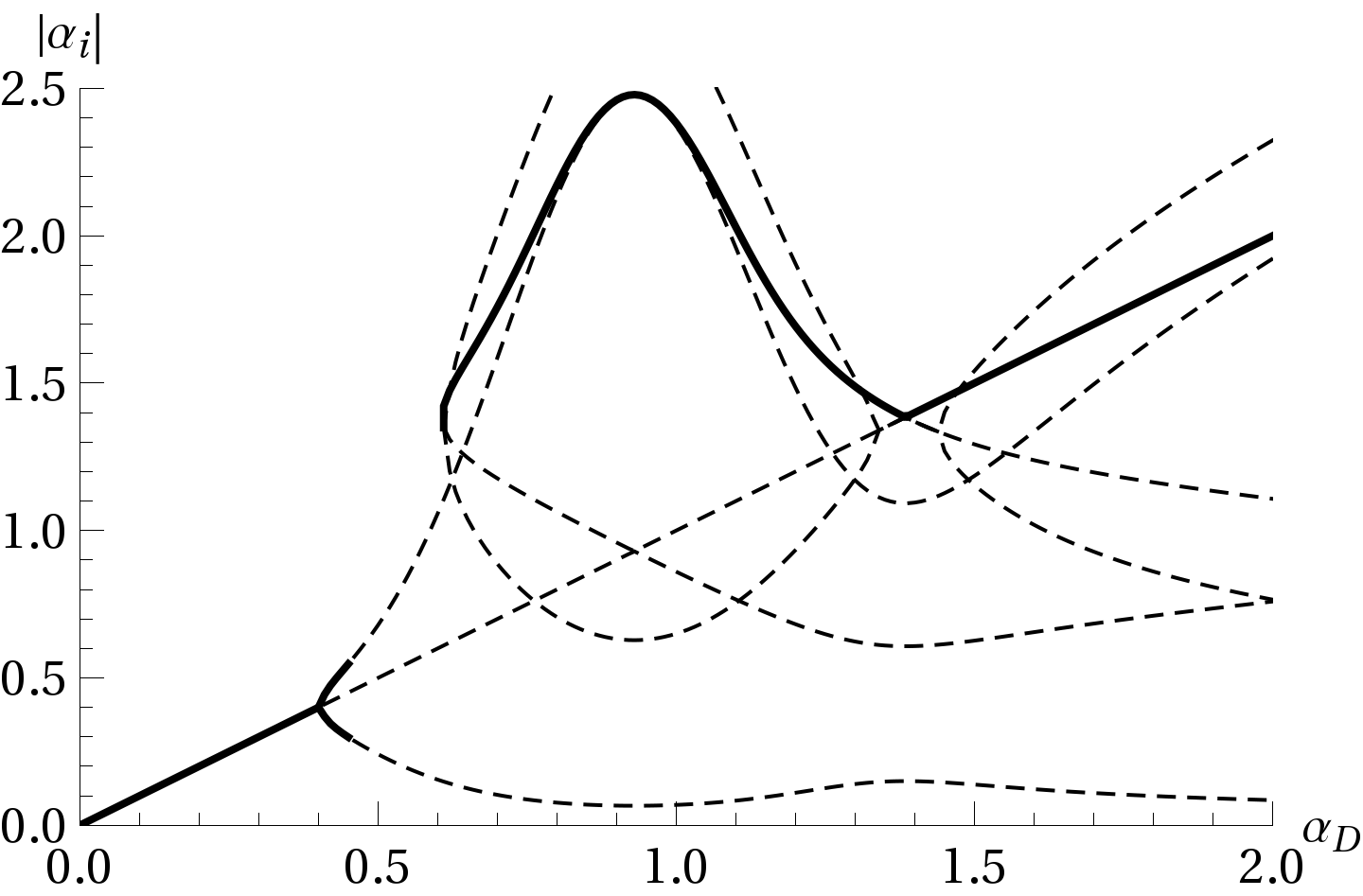}
\caption{\label{fig:alpha} Mean intracavity field $|\alpha_i|$ as a function of $\alpha_D$. All parameters are as in Fig.~\ref{fig:etavalpha}. Stable solutions are solid and unstable are dashed. The dark sine mode ansatz indeed produces a solution over the entire range. However, it is only unique and stable for very feeble light fields. As the intensity is increased, the solution bifurcates and becomes unstable, returning to stability for sufficiently high intracavity field strength (though it is accompanied by additional unstable solutions).}
\end{figure}
Recall from Eqs.~(\ref{xcclass})-(\ref{defineZ}) that $\bar X_c$ has a nontrivial implicit dependence on $\alpha_D$ (proportional to $\alpha_D^2$ for small $\alpha_D$ and proportional to $\alpha_D^{-6}$ for large $\alpha_D$). Once we begin the discussion of the quantum fluctuations, we will treat $\alpha_D$ as the free parameter rather than $\eta$, because it allows the quantum equations of motion to be put forth in closed form. That such a substitution is possible without requiring an inordinately large pump intensity $\eta$ is shown in Fig.~\ref{fig:etavalpha}, a plot of the relationship between the desired intracavity field $\alpha_D$ and the imaginary part of the required $\eta$. This replacement can therefore allow $\alpha_D$ to be interpreted as the {\em predicted} intracavity field strength for a given pumping strength under the assumption that there are no atoms excited to the sine side mode.

Under the restrictions (\ref{etatoalpha}), a bifurcation diagram of the steady state solutions of the mean intracavity field strengths $|\alpha_i|$ as a function of $\alpha_D$ are shown in Fig.~\ref{fig:alpha}. In general, the system has up to six steady-state solutions, with up to two being stable. For small $\alpha_D$, we have $\alpha_1 = \alpha_2=\alpha_D$, but as $\alpha_D$ is increased the system undergoes a bifurcation, with $\alpha_1 \neq \alpha_2$. This occurs at $\alpha_D \approx .4$ for our parameters. This is followed by a small window with no stable solution (between about $\alpha_D = .45$ and .6), at which point, there is again a stable symmetric solution where the $\alpha_i$ are larger than might be expected. Presumably, the symmetry breaking creates something of a positive feedback loop: a few atoms are excited into the sine mode, which shifts the phase of the counterpropagating light fields, causing them to interfere with each other less effectively, thereby increasing the net number of photons in the cavity, causing more atoms to be excited. Lastly, for much larger light fields, the pumping dominates the atoms' symmetry breaking ability and the dark sine mode becomes stable again.
\begin{figure}[t]
\includegraphics[width=0.45 \textwidth]{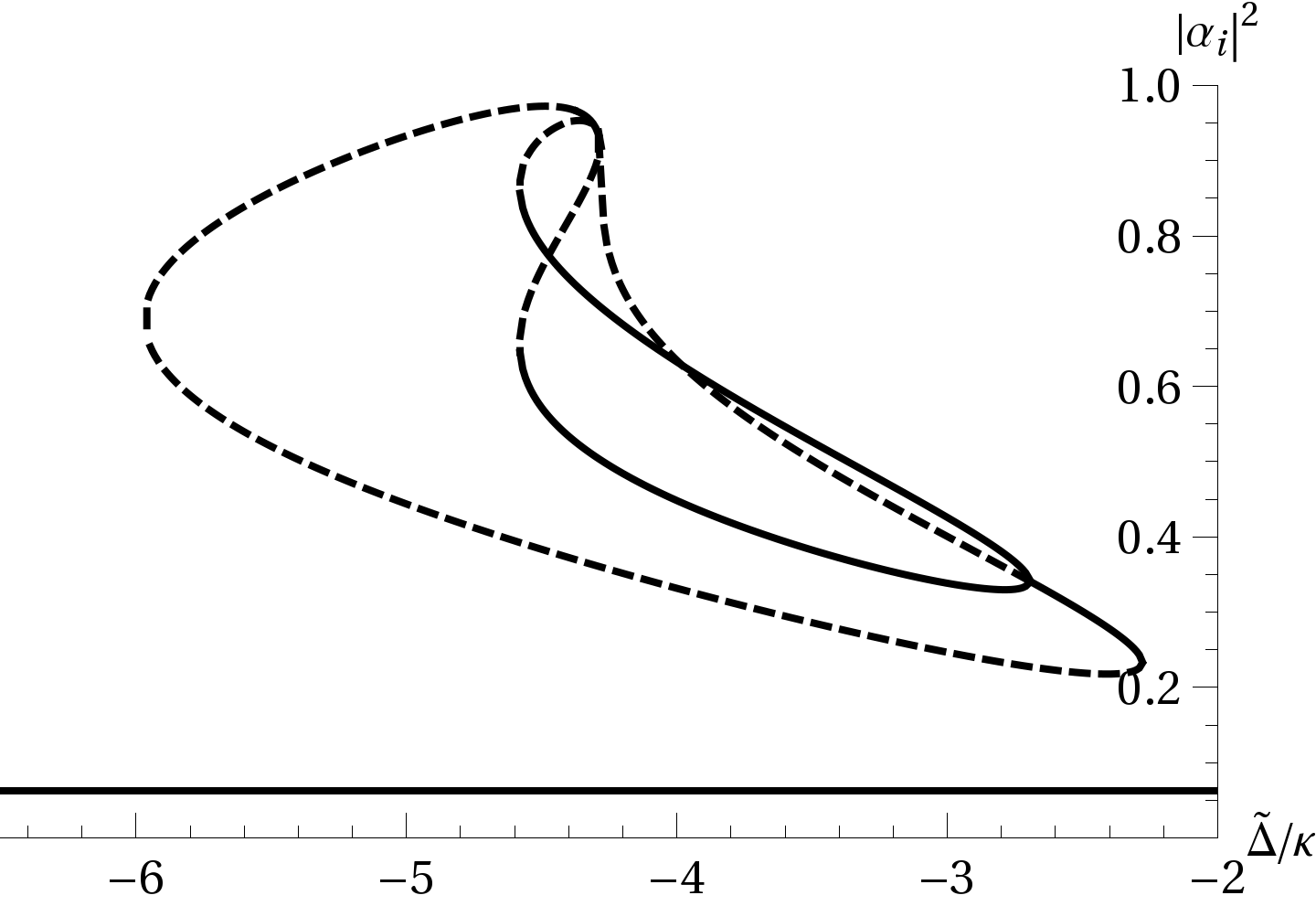}
\caption{\label{fig:delta1} Mean intracavity photon number $|\alpha_i|^2$ as a function of $\tilde\Delta$. All parameters are as in Fig.~\ref{fig:bifurc}, except rather than a fixed $\eta$, we take $\alpha_D=0.25$. Stable solutions are solid and unstable are dashed.}
\includegraphics[width=0.45 \textwidth]{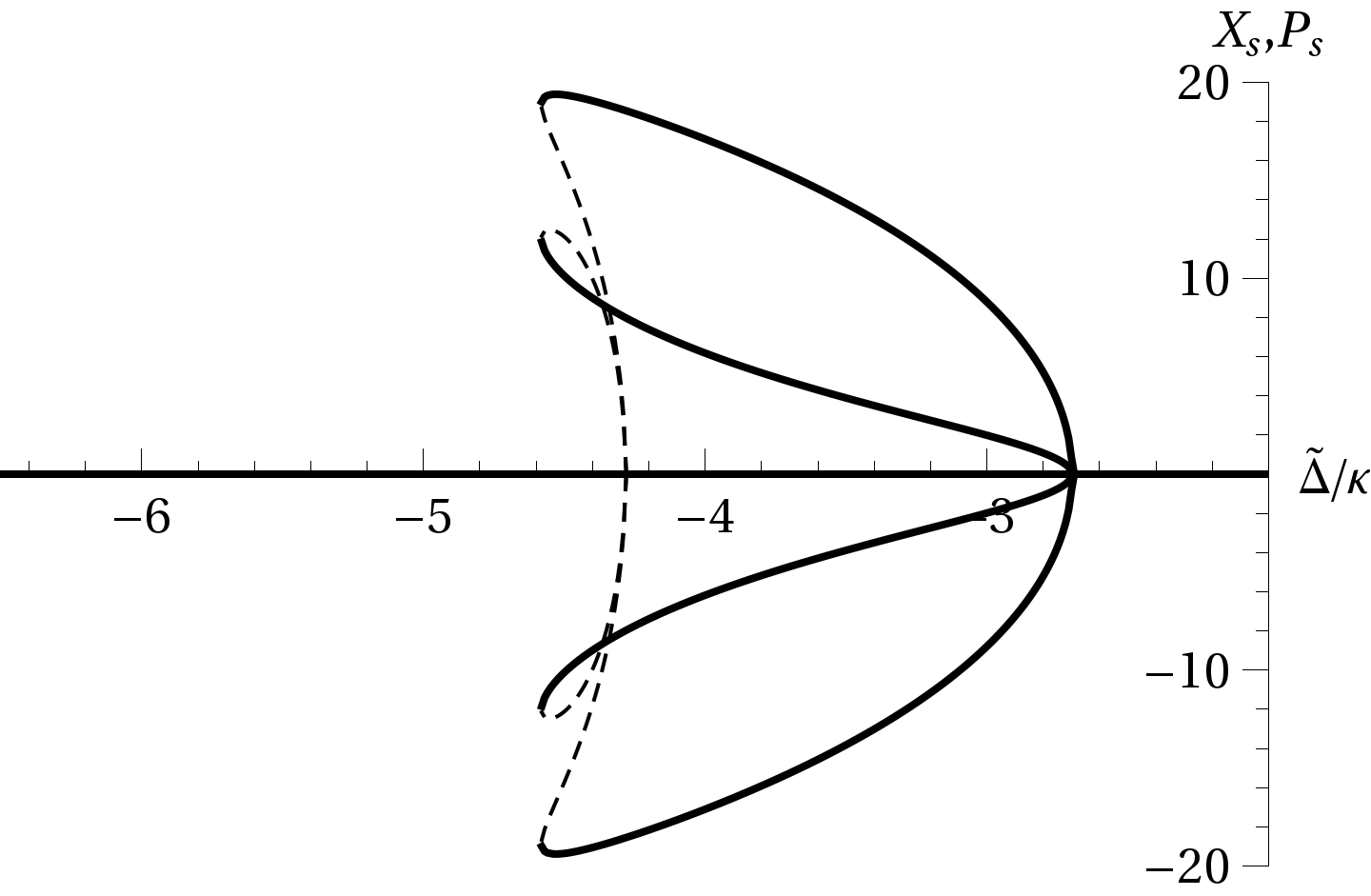}
\caption{\label{fig:delta2} Mean sine mode position and momentum as a function of $\tilde\Delta$. All parameters as in Fig.~\ref{fig:delta1}. The outer branches are position solutions and the inner are momentum.}
\end{figure}
\begin{figure}[t]
\includegraphics[width=0.45 \textwidth]{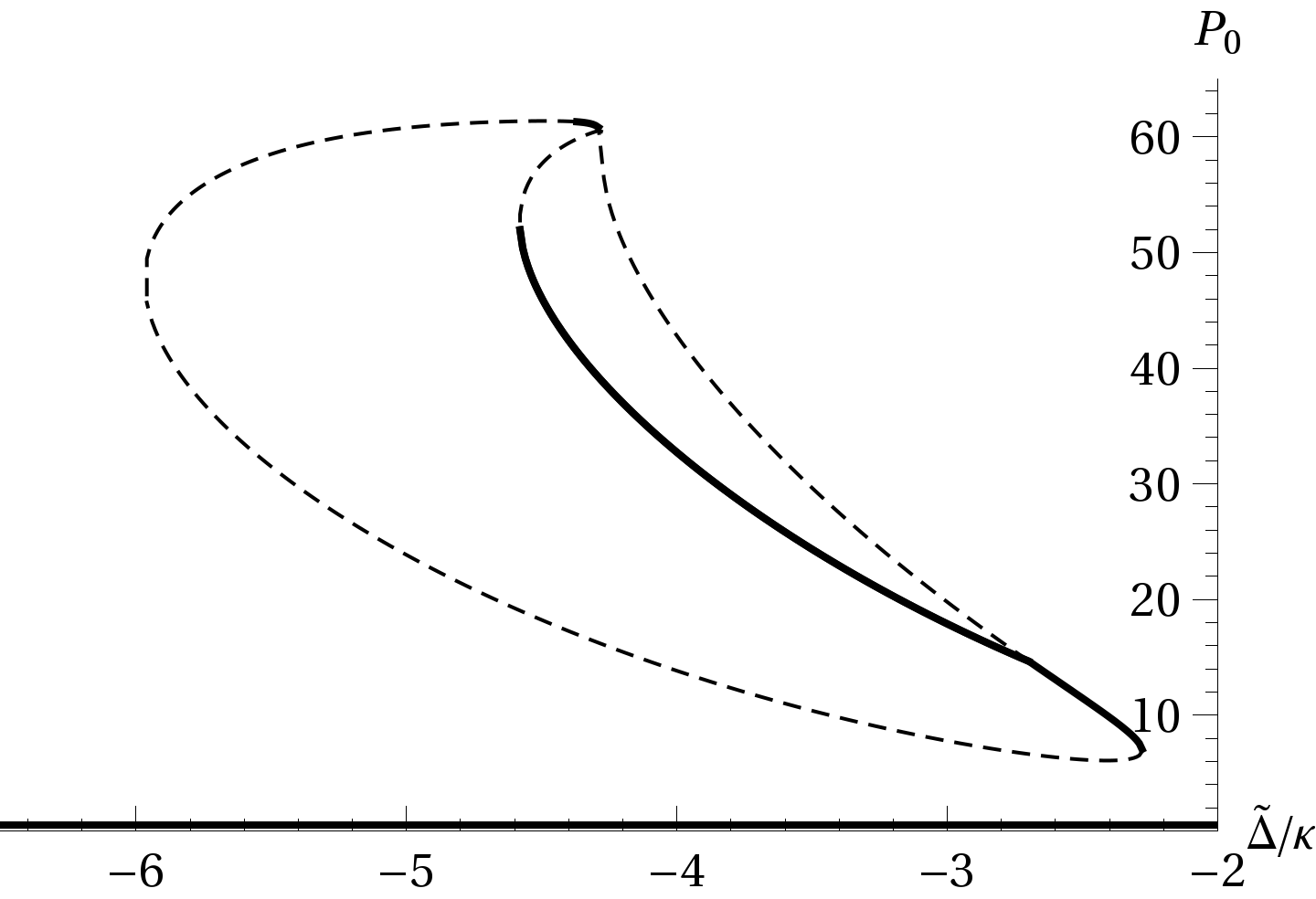}
\caption{\label{fig:delta3} Mean $\bar P_0$ as a function of $\tilde\Delta$. All parameters are as in Fig.~\ref{fig:delta1}.}
\end{figure}
In addition, the character of the bifurcation diagrams as functions of $\tilde\Delta$ will change because of (\ref{etatoalpha}), that is, because we use a complex (rather than purely real as in Fig.~\ref{fig:bifurc}) $\eta$ whose imaginary part is itself a linear function of $\tilde\Delta$. These diagrams are shown in Figs.~\ref{fig:delta1}--\ref{fig:delta3}. The multistable behaviors in particular are richer than before. The dark sine mode remains stable over the entire range of $\tilde\Delta$ (straight solid line $|\alpha_i|^2=.0625=.25^2$ at the bottom of Fig.~\ref{fig:delta1}), yet it is augmented by an isola consisting of additional steady-state solutions with macroscopic side mode occupancies. In particular, $\bar N_s$ can reach approximately 250, almost 3\% of the total number of atoms (see Fig.~\ref{fig:delta2}). The stability and symmetry of these solutions varies greatly depending on the detuning. We will see later that quantum fluctuations may affect tunneling between these stable solution branches. We remark that for the range of detunings where the isola has a single stable solution (between $\tilde\Delta\approx-2.2\kappa$ and $\tilde\Delta\approx-2.7\kappa$) the light mode has an additional symmetric stable solution, but the side mode still remains dark. It becomes macroscopically occupied once the symmetry of the optical isola is broken. Lastly, we examine the effect of allowing depletion of the zero-momentum condensate mode; that is, we look at the steady-state values $\bar X_0$ and $\bar P_0$. From Eq.~\eqref{c0}, we have $\bar c_0 = \sqrt{N}+\bar X_0/\sqrt{2}+i\bar P_0/\sqrt{2}$. We find that $\bar X_0$ (not plotted here) is negative in sign and quite a bit smaller in magnitude than $\sqrt{N}$; considering only this fact we might wonder if the evolution of the zero-momentum mode is really worth examination. However, when the system's symmetry is broken and the sine mode acquires a finite occupancy, $\bar P_0$ can actually be fairly significant compared to the relevant scale, $\sqrt{2N}\approx 134$ (it should be noted that $\bar X_0$, though small in magnitude, still reduces the real part of $\bar c_0$ enough such that $|\bar c_0|^2 \le N$; the condensate cannot ``grow'' from this treatment). This result strongly suggests that the replacement of $\hat c_0$ by $\sqrt{N}$ is not always the optimal ansatz to make; here it seems that $\hat c_0$ should be a dynamical, rather than static, quantity. To put this another way, we expect the classical steady-state solutions of the equations of motion to be comprised of coherent states of the light and matter fields, including the zero-momentum mode. When the sine mode attains a finite occupancy, its presence can cause a major shift in the quantum state of the zero-momentum mode beyond just the removal of a few atoms.

\section{Quantum Effects}

\subsection{Linearized quantum equations of motion}
We now turn to a discussion of the impact of small quantum fluctuations on the dynamics of the system. To proceed, and armed with our detailed  results on the mean-field behavior in hand, we include quantum fluctuations in the original equations of motion, assuming that these fluctuations remain sufficiently small that their effects may adequately be described by linearized equations of motion. Specifically we introduce the fluctuations in the familiar fashion via small quantum corrections to the mean-field solutions,
\begin{equation}
\hat X_0 \rightarrow \bar X_0 + \hat x_0,
\end{equation}
and so forth, and linearize the Heisenberg-Langevin equations of motion in these fluctuations. This process is justified as long as the quantum fluctuations are small compared to the classical means (or in the case of the sine mode, with its zero mean, as long as the quadratic terms are smaller than the linear ones). But we remark again that because of the instabilities demonstrated here and in Ref.~\cite{chen2010} this will not always be the case, so some care must be taken when selecting values for the parameters.

This linearization procedure yields the ten coupled operator equations of motion
\beq
\doh x_i &=& -\kappa\hat x_i - \tilde\Delta\hat p_i + \Omega_B\hat p_j \nonumber \\
&+&(-1)^i(\chi_0\hat x_s+ \wp_0\hat p_s)+ \hat\xi_{xi},\nonumber\\
\doh p_i &=& -\kappa\hat p_i + \tilde\Delta\hat x_i-\Omega_B\hat x_j - \chi_c\hat x_0 \nonumber \\
&-&\chi_0\hat x_c - \wp_c\hat p_0 - \wp_0\hat p_c + \hat\xi_{pi},\nonumber\\
\doh x_0 &=& -\gamma\hat x_0 + \Omega\hat p_c+\wp_c(\hat x_1+\hat x_2)+\hat\xi_{x0},\nonumber\\
\doh p_0 &=& -\gamma\hat p_0 - \Omega\hat x_c-\chi_c(\hat x_1+\hat x_2)+\hat\xi_{p0},\nonumber\\
\doh x_c &=& -\gamma\hat x_c +\omega_r\hat p_c + \Omega\hat p_0 + \wp_0(\hat x_1 + \hat x_2)+\hat\xi_{xc},\nonumber\\
\doh p_c &=& -\gamma\hat p_c -\omega_r\hat x_c - \Omega\hat x_0 - \chi_0(\hat x_1 + \hat x_2)+\hat\xi_{pc},\nonumber\\
\doh x_s &=& -\gamma\hat x_s +\omega_r\hat p_s + \wp_0(\hat p_2 - \hat p_1)+\hat\xi_{xs},\nonumber\\
\label{QEOM}\doh p_s &=& -\gamma\hat p_s -\omega_r\hat x_s - \chi_0(\hat p_2 - \hat p_1)+\hat\xi_{ps},
\eeq
where $i= \{1,2\}$, $j = 3 - i$, and
\beq
\Omega_B &=& \frac{\Omega_0}{\sqrt 2}\left[(\sqrt{2N}+\bar X_0)\bar X_c + \bar P_0\bar P_c\right],\nonumber\\
\chi_0 &=& \Omega_0\alpha_D(\sqrt{2N}+\bar X_0),\nonumber\\
\chi_c &=& \Omega_0\alpha_D\bar X_c,\nonumber\\
\wp_0 &=& \Omega_0\alpha_D\bar P_0,\nonumber\\
\wp_c &=& \Omega_0\alpha_D\bar P_c.
\eeq
The first two of Eqs.~(\ref{QEOM}) describe the fluctuations of the light field, and the last six the matter-wave fluctuations in the zero-momentum component and in the sine and cosine side modes. The terms proportional to $\Omega_B$ describe  Bragg scattering between the two counter-propagating optical fields due to the material grating formed by the zero-momentum matter wave and the cosine mode. The coupling between the light and matter operators is determined by the constants $\chi_0, \chi_c, \wp_0$ and $\wp_c$, which act as small perturbations that couple the evolution of the four light and six matter operators. Note that there is no coupling dependent on the occupancy of the sine side mode at the level of these equations of motion. The coupling coefficients involve only the classical, mean optical and matter wave fields. Thus, the coupling to the sine side mode only occurs indirectly via quantum fluctuations in the symmetric driving situation considered here.

While the exact eigenvalues and eigenvectors of the 10$\times$10 matrix defined by these equations cannot be found explicitly in closed form, standard perturbation theory states that the eigenvalues associated to the {\em uncoupled} optical and matter blocks of equations, e.g. for $\chi_{0,c}=\wp_{0,c}=0$, match those of the coupled system up to second order in the perturbation. Numerical testing confirms that the values indeed remain close. These eigenvalues are
\beq
\lambda_{Lc} &=& -\kappa\pm i(\tilde\Delta+\Omega_B),\nonumber\\
\lambda_{Ls} &=& -\kappa\pm i(\tilde\Delta-\Omega_B),\nonumber\\
\lambda_{Mc} &=& -\gamma\pm i\sqrt{\Omega^2+\frac{\omega_r}{2}\left(\omega_r\pm\sqrt{4\Omega^2+\omega_r^2}\right)},\nonumber\\
\lambda_{Ms} &=& -\gamma\pm i\omega_r.
\eeq
The reason for this nomenclature is clear when one considers the corresponding eigenvectors (normal modes of evolution). While their explicit expressions are not possible to write in closed form and are exceedingly unwieldy even when only expanded to first order in the perturbation, a qualitative inspection yields some useful information. The two pairs of eigenvalues $\lambda_{Lc}$ and $\lambda_{Ls}$ correspond to light-dominated evolution with a small mixture of cosine and zero-momentum state matter modes, in the first case, and of the sine matter mode, in the second case. As one might anticipate, the oscillation frequencies of these optical modes have shifts of opposite sign depending on whether they are coupled to the symmetric or antisymmetric matter grating. The four $\lambda_{Mc}$ eigenvalues correspond to normal modes dominated by the cosine mode and zero-momentum mode (in unequal proportion) and coupled to the light fields. The $\lambda_{Ms}$ values correspond to sine-dominated normal modes coupled to the light fields. The sine-dominated normal modes only contain a tiny contribution from the cosine mode and zero-momentum modes and vice-versa, a direct consequence of the dark-state nature of the sine mode at the classical, mean-field level. This explains why the sine normal mode's oscillation frequency is just $\omega_r$: to lowest order, it is decoupled from the other normal modes of the system. Since $\gamma$ is typically the smallest dimensional parameter, as one increases the coupling between the light and matter fields, for example by increasing $\alpha_D$, the real part of $\lambda_{Mc}$ can become positive even though the change in $|\lambda_{Mc}|$ is relatively small. This leads to the instability in the dark sine mode solution seen in Fig.~\ref{fig:alpha}. Because $\kappa \gg \gamma$ the optical transients die out rapidly and the light fields follow adiabatically the matter wave fields, with noise- and interaction-dominated fluctuations.

\subsection{A formal parenthesis}
To proceed further we open a small parenthesis to introduce a somewhat formal result that will prove useful in the analysis of higher-order correlation functions of the quantized atomic and optical fields, in particular the correlations of various orders of the matter-wave and light modes, as well as the cross-correlations between the matter and light fields.This formal development extends much of the machinery familiar from systems with a single damped operator to deal with multiple noise sources with distinct characteristics.

Consider a quantum system described, in the Heisenberg picture, by a set of $d$ operators $\hat O_k$ that comprise a $d$-dimensional vector operator $\hvc O$. Let $\hvc O$ evolve according to a linear (or linearized) system of $d$ coupled Heisenberg-Langevin equations,
\beq
\ddt\hvc O(t) &=& {\mathbf W}\hvc O(t) + \hvc \xi(t),
\eeq
where $\mathbf W$ is a $d \times d$ matrix with c-number coefficients and $\hvc \xi$ are noise operators with 0 mean (i.e., any net input to the system has already been absorbed into the equations of motion). For convenience we have merged the usual factors of $\sqrt{2\kappa}$ needed to preserve commutation relations into $\hvc \xi$, since they may vary for each operator.

These equations can be integrated in a straightforward fashion, yielding
\begin{equation}
\hvc O(t) = e^{{\mathbf W}t}\hvc O(0)+\int_0^t e^{{\mathbf W}(t-u)}\hvc \xi(u)du.
\end{equation}
For each $i,j\le d$, and at all times $s,t>0$ we have
\beq
\langle \hat O_i(0)\hat \xi_j(t)\rangle &=& 0,\\
\langle \hat \xi_i(s)\hat \xi_j(t)\rangle &=& {\mathbf N}_{ij}\delta(s-t),
\eeq
where the first condition is satisfied axiomatically, and the second one holds for white noise sources, an approximation that should be adequate for the system under study. The expectation values of the operators alone are just
\begin{equation}
\langle\hvc O(t)\rangle = e^{{\mathbf W}t}\langle \hvc O(0)\rangle,
\end{equation}
since $\langle \hvc \xi(t)\rangle = 0$.

Under these conditions we can obtain the correlation matrix
\beq
\langle\hvc O(s)\otimes\hvc O(t)\rangle&=&e^{{\mathbf W}s}\langle\hvc O(0)\otimes\hvc O(0)\rangle e^{{\mathbf W}^T t}\nonumber\\
&+&\int_0^{{\rm min}(s,t)}\hspace{-12 pt}e^{{\mathbf W}(s-u)}{\mathbf N}e^{{\mathbf W}^T(t-u)}du,
\eeq
which lets us compute all quadratic correlations at all times,
e.g. $\langle \hat x_1(s) \hat p_c(t)\rangle$.

For the rest of the paper, we shall work in the long-time limit, in which the transient behavior of the operators has decayed to 0. In physical terms, for the model in question, this corresponds to times $t \gg 1/\gamma$, which are experimentally accessible for long-lived BECs. In this limit, the initial values have decayed to irrelevance, and the correlations are simply
\beq
\langle\hvc O(s)\otimes\hvc O(t)\rangle&=&\int_0^{{\rm min}(s,t)}\hspace{-12 pt}e^{{\mathbf W}(s-u)}{\mathbf N}e^{{\mathbf W}^T(t-u)}du.
\eeq
This is the central result of this section. For a Gaussian process we can easily determine higher order correlations as well: all three-operator correlators are 0, and the four operator correlators are given by
\beq
& &\hspace{-48 pt}\langle\hvc O(s)\otimes\hvc O(t)\otimes\hvc O(u)\otimes\hvc O(v)\rangle = \nonumber\\
& &\langle\hvc O(s)\otimes\hvc O(t)\rangle\otimes\langle\hvc O(u)\otimes\hvc O(v)\rangle\nonumber\\
&+&\langle\hvc O(s)\otimes\hvc O(u)\rangle\otimes\langle\hvc O(t)\otimes\hvc O(v)\rangle\nonumber\\
&+&\langle\hvc O(s)\otimes\hvc O(v)\rangle\otimes\langle\hvc O(t)\otimes\hvc O(u)\rangle.
\eeq

Now, we apply this technique to our model. The operators $\hvc O$ are the $\hat x$'s and $\hat p$'s, with the
coefficients $\mathbf W$ given by (\ref{QEOM}). Lastly, we determine the $\xi_{x,p}$, and hence,
$\mathbf N$ from the following relations:
\beq
\langle \hat \xi_{ai}(s)\hat \xi_{ai}^\dagger(t)\rangle &=& 2\kappa\delta(s-t)(N_i^{th}+1),\\
\langle \hat \xi_{ai}^\dagger(s)\hat \xi_{ai}(t)\rangle &=& 2\kappa\delta(s-t)N_i^{th},\\
\langle \hat \xi_{cI}(s)\hat \xi_{cI}^\dagger(t)\rangle &=& 2\gamma\delta(s-t)(N_I^{th}+1),\\
\langle \hat \xi_{cI}^\dagger(s)\hat \xi_{cI}(t)\rangle &=& 2\gamma\delta(s-t)N_I^{th},
\eeq
where $i = \{1,2\}$, $I=\{0,c,s\}$, the $N^{th}$'s are thermal noise occupancies of the baths near the characteristic frequencies of the system as given by Bose-Einstein statistics, and all other quadratic noise correlations are 0. The noise matrix $\mathbf N$ for the position and momentum operators thus has the form of 5 $2\times 2$ block matrices, each with $2 N^{th}+1$ for the on-diagonal entries and $\pm i$ for the off-diagonal entries with all 4 entries multiplied by $\gamma$ or $\kappa$ as appropriate. Because we are dealing with optical photons and a BEC at a temperature of at most a few $\mu$K, going forward we take all $N_{th}\rightarrow 0$, or, equivalently, we take the bath temperatures to be 0. We find that typically increasing the $N_{th}$'s just adds directly to the occupancy of the corresponding fields.

\subsection{Second-order correlations and quantum occupancies}
With these formal results at hand, and working with a set of parameters that are a combination of those in Figs.~\ref{fig:alpha} -- \ref{fig:delta3} and with $\tilde\Delta = -1. \times \kappa$, $\alpha_D=0.25$, we are now in a position to explore a few results for the cross operator correlations, before looking at the quantum-fluctuation-augmented occupancy of the side modes. Further calculations not presented here have shown that the results obtained for these particular parameter values are fairly typical of the monostable regime in which we are interested.

As expected, the fluctuations in the zero-momentum and cosine modes are virtually uncorrelated with those of the sine mode, $\langle(\hat x,\hat p)_{0,c}(\hat x,\hat p)_s\rangle\approx 0$, but are slightly correlated with each other (e.g. $\langle \hat x_0 \hat x_c \rangle \approx 0.022$ -- for comparison, $\bar X_0\bar X_c \approx 0.11$, so the classical mean-field correlation is a more significant contribution). By far the largest correlation between distinct matter and/or light fields, however, is the one that confirms our intuition, namely, the sine mode's fluctuations are very strongly correlated to those of the light field (e.g. $\langle\hat x_i\hat x_s\rangle\approx(-1)^i0.32, i =1,2$). This shows that, indeed, the occupation of the sine side mode is driven by and interdependent with the fluctuations in the light fields.

\begin{figure}[t]
\includegraphics[width=0.45 \textwidth]{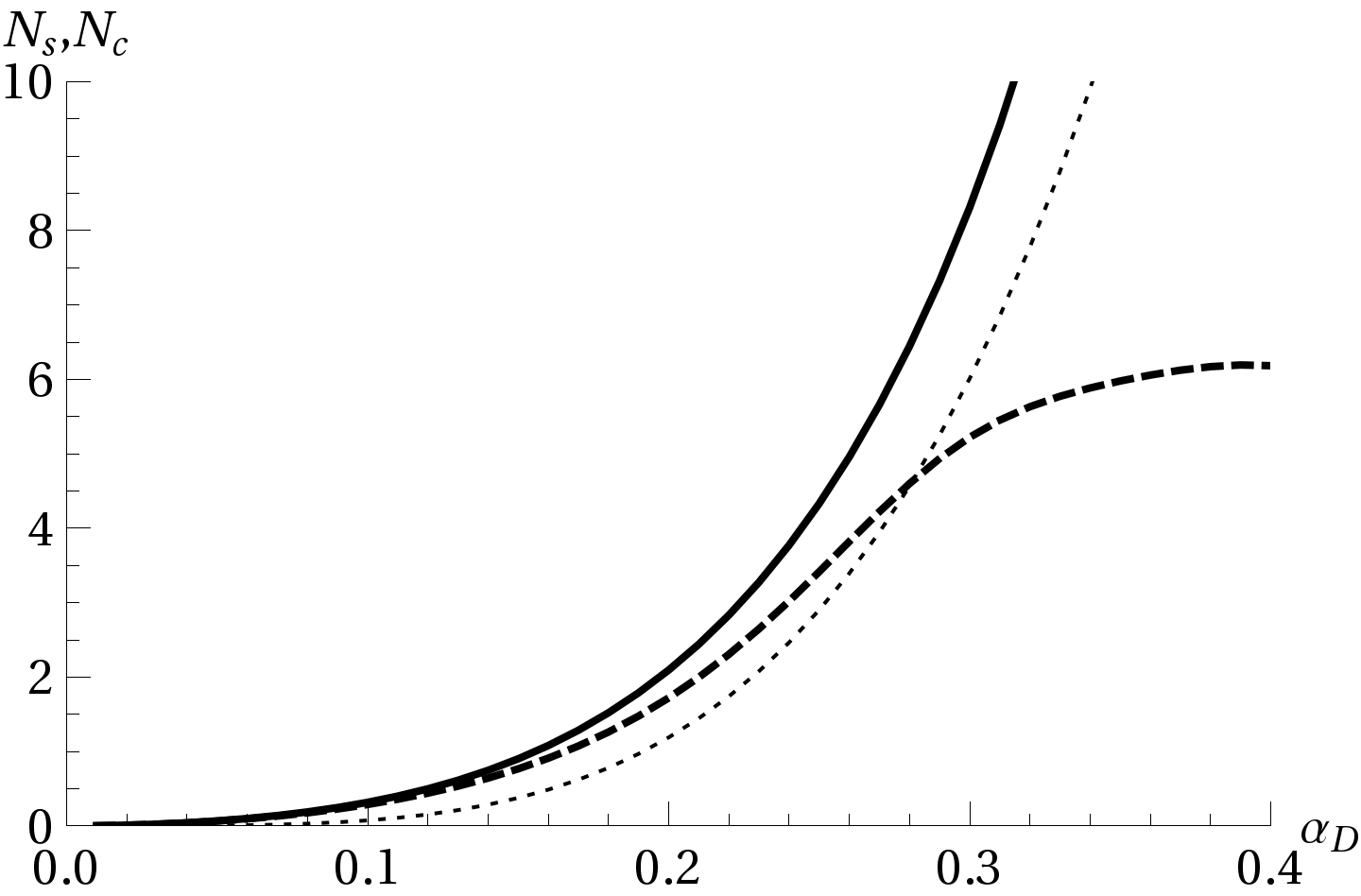}
\caption{\label{fig:nvalpha} Side mode occupancies $\langle N_c \rangle,\langle N_s \rangle$ as functions of $\alpha_D$. Parameters as in Fig.~\ref{fig:etavalpha}. The cosine mode (solid) has a larger occupancy than that of the sine mode (dashed). For reference, the classical mean $\bar N_c$ is plotted as well (dotted line).}
\includegraphics[width=0.45 \textwidth]{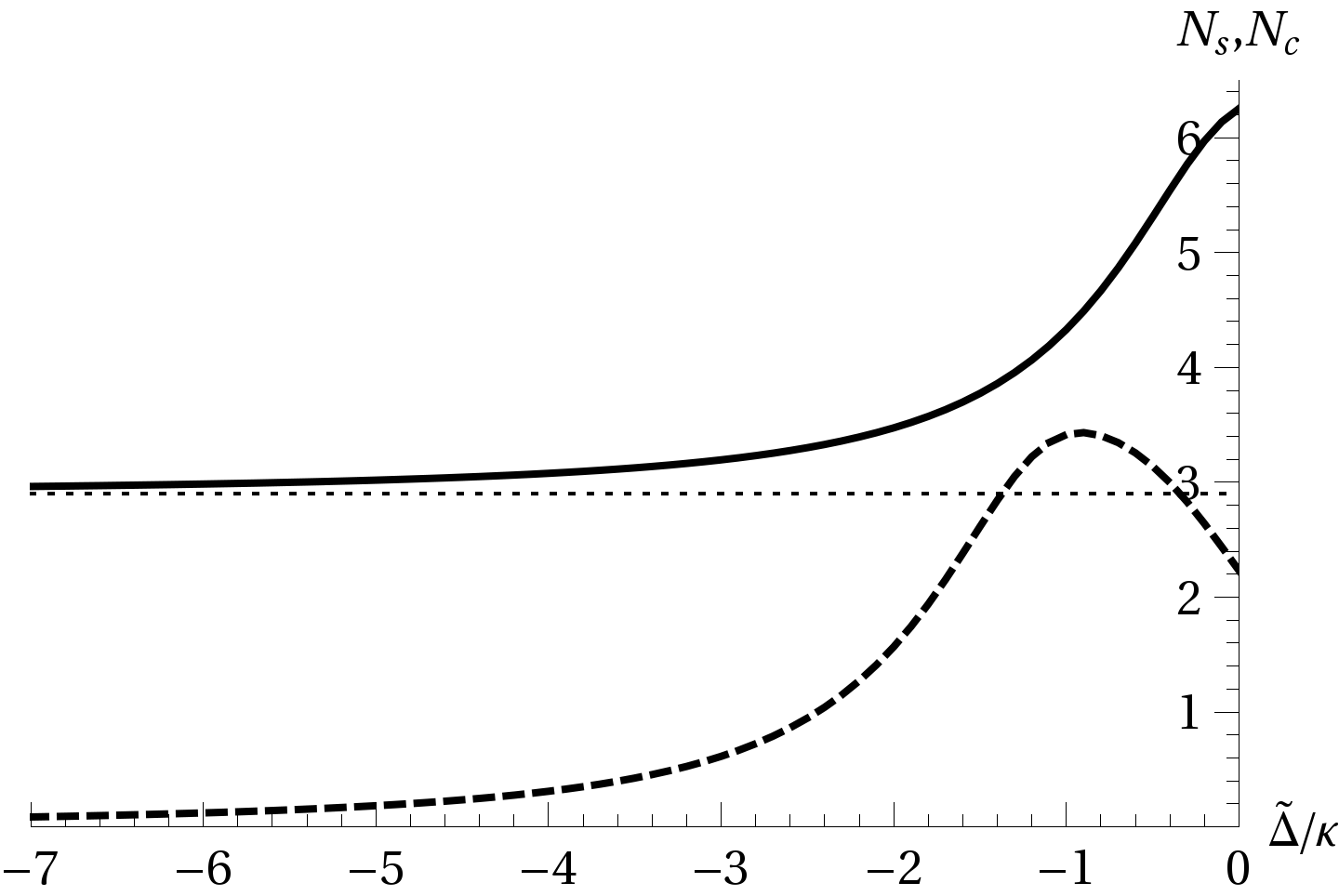}
\caption{\label{fig:nvdelta} Side mode occupancies $\langle N_c \rangle,\langle N_s \rangle$ as functions of $\tilde\Delta$. Parameters as in Fig.~\ref{fig:delta1}, but the range of $\tilde\Delta$ has been extended slightly. The cosine mode (solid) again has a larger occupancy than the sine mode (dashed). For reference, the classical mean $\bar N_c$ is plotted as well (dotted); $\bar X_c$ and $\bar P_c$ do not depend on $\tilde\Delta$ so it is constant.}
\end{figure}
We also consider the occupancy of the side modes as functions of the most easily tunable parameters $\tilde\Delta$ and $\alpha_D$. Keeping in mind that all single operator expectation values such as $\langle \hat x_i\rangle$ decay to 0, we have
\beq
\langle \hat X_i^2\rangle &=& \bar X_i^2 + \langle \hat x_i^2\rangle,
\eeq
etc. When evaluating these quantities we must be careful to avoid those regions in parameter space where the dark sine mode steady state solution is unstable, in particular, we need $\alpha_D<0.4$. The results are shown in Figs.~\ref{fig:nvalpha} and \ref{fig:nvdelta}. In the former, we see a noticeable occupation of the sine mode before the classical bifurcation. This may be sufficient to shift the bifurcation point to a lower value of $\alpha_D$. This possibility is corroborated by the observed shift to the left in the plot of $\langle N_c \rangle$ versus $\bar N_c$; that is, when quantum fluctuations are included, the cosine mode behaves as if $\alpha_D$ were slightly larger. On the other hand, a somewhat different behavior is seen in the latter plot. As $\tilde\Delta$ decreases from 0 and approaches the bifurcation seen above in Figs.~\ref{fig:delta1}--\ref{fig:delta3}, the sine mode initially starts to increase in occupancy, but then its (and the cosine mode's) quantum fluctuations are suppressed as $|\tilde\Delta|$ increases further. Nevertheless, as we see below, the variance in the sine mode's occupation is so large that the quantum fluctuations may still influence the character of the system's behavior in the case $-2.5\kappa \stackrel{<}{\sim} \tilde\Delta \stackrel{<}{\sim} -1.0\kappa$.

\subsection{Variance in side mode occupancy}
Because the system is coupled to a zero temperature bath, we compare the variance
\beq
\label{sigma}\sigma_I^2 &=& \langle\hat N_I^2\rangle-\langle\hat N_I\rangle^2
\eeq
to that of a bosonic system in thermal equilibrium, in which case
\beq
\label{sigmaTH}\sigma_{I,\rm TH}^2 &=& \langle\hat N_I\rangle^2+\langle\hat N_I\rangle,
\eeq
\begin{figure}[b]
\includegraphics[width=0.45 \textwidth]{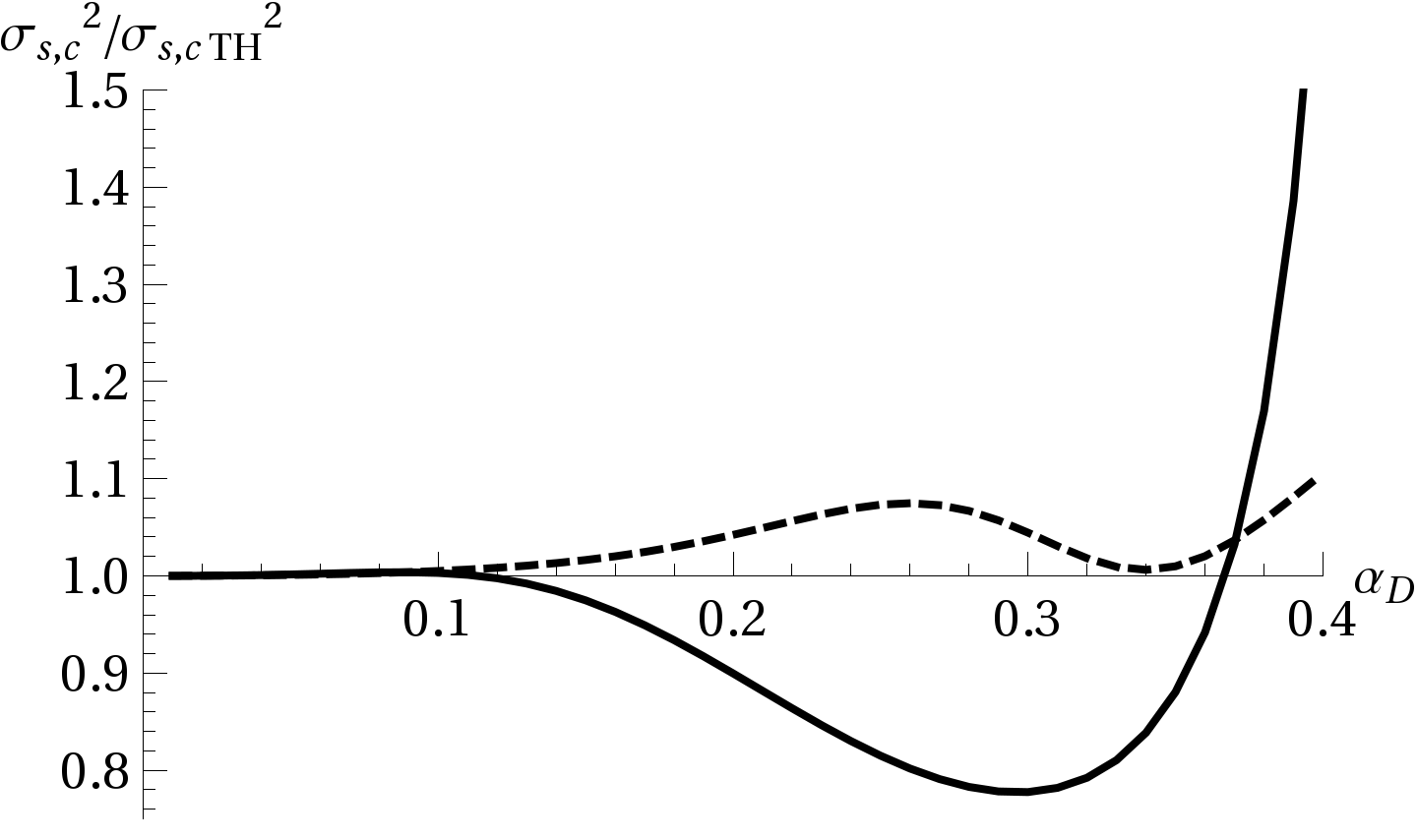}
\caption{\label{fig:sigvalpha}Variance of side mode occupancy compared to thermal variance as functions of $\alpha_D$. Parameters as Fig.~\ref{fig:nvalpha}. Cosine mode is solid and sine mode is dashed.}
\end{figure}
specifically taking the ratio of (\ref{sigma}) to (\ref{sigmaTH}). A value less than one indicates sub-thermal statistics, as would be the case when there is a significant classical mean and/or quantum fluctuations are suppressed. On the other hand, a ratio greater than one indicates significant fluctuations and a matter or light field driven out of thermal equilibrium. These ratios are computed and plotted in Figs.~\ref{fig:sigvalpha} and \ref{fig:sigvdelta} as functions of $\alpha_D$ and $\tilde\Delta$, respectively. In the former, for weak mean intracavity light field $\alpha_D$, both modes' statistics are thermal in nature. As the applied field is increased, the sine mode is perturbed to slightly higher variance, whereas the cosine mode at first exhibits less than thermal variance, as the classical mean field contribution grows. However, the quantum fluctuations then take over and its variance grows quickly as $\alpha_D$ approaches the bifurcation at a value of roughly 0.4. We should however take this result with a grain of salt since it is at this point that the quantum fluctuation contribution to $\langle \hat N_c\rangle$ exceeds the mean contribution $\bar N_c$, thus endangering the validity of the linearized treatment. Still, it should of course be expected that increased fluctuations in the cosine mode significantly alter the existence and stability of steady state solutions in this critical region.

In the second of these figures we plot the variances as a function of the effective detuning $\tilde\Delta$. In this case, except for detunings very near 0, the cosine mode is almost completely classical, as is the sine mode for sufficiently negative values of $\tilde\Delta$. This implies that, if a zero-momentum condensate is formed and allowed to evolve for the parameters given and a $\tilde\Delta$ of less than $-2.5\kappa$ or so, and if it reaches the dark sine mode steady state, it is quite likely to remain there indefinitely, as quantum fluctuations are strongly suppressed. But for less negative values of $\tilde\Delta$, the sine mode fluctuations are significant. It may be possible that these fluctuations ``anticipate'' the classical bifurcation nearby in parameter space, or even that they allow the new stable solutions to appear for larger values of the detuning than they would otherwise. To test this would likely require simulation of the full nonlinear quantum evolution of the system.

\begin{figure}[t]
\includegraphics[width=0.45 \textwidth]{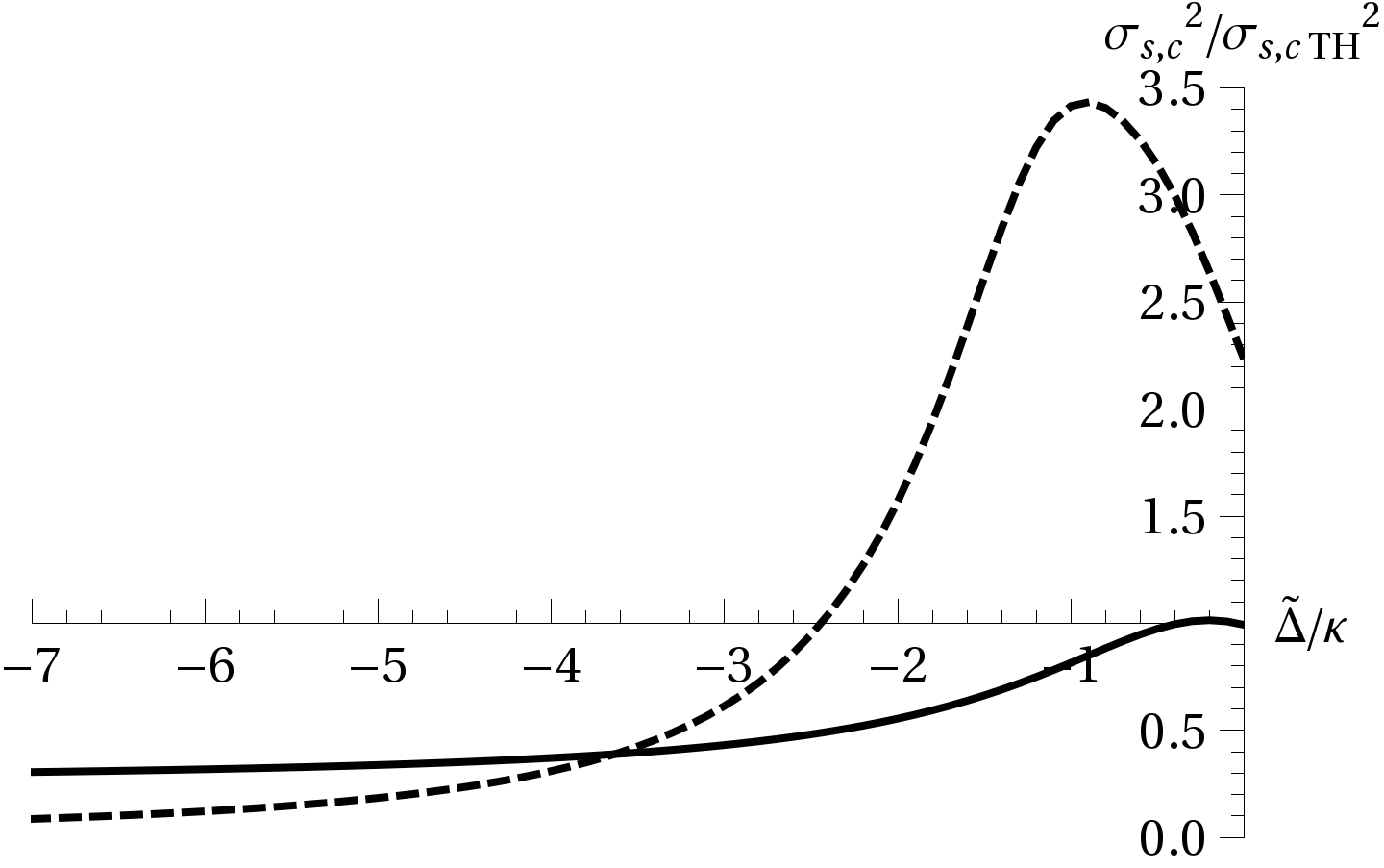}
\caption{\label{fig:sigvdelta}Variance of side mode occupancy compared to thermal variance as functions of $\tilde\Delta$. Parameters as Fig.~\ref{fig:nvdelta}. Cosine mode is solid and sine mode is dashed.}
\end{figure}

\section{Discussion and Conclusions}

By analyzing a detailed model including two counter-propagating light fields and three matter fields, we are able to find a region in parameter space, with experimentally accessible values, where the system's behavior differs significantly from that of a BEC in a Fabry-P\'{e}rot cavity, and also where quantum corrections become significant. The classical dynamics are rich, and near bifurcation points in the mean-field classical system, the quantum fluctuations also have intriguing properties. They appear strong enough to shift or perturb the dynamical bifurcation points.

This system's dynamics are richer than the typical optomechanical system, and they may be exploited in the future to investigate numerous non-classical effects. For instance, because of the strong cross-correlation between the light's and the sine mode's fluctuations, it should be possible, by measuring the output light fields, to monitor the sine mode as it tunnels between different possible steady state branches. This would contrast with the tunneling suppression seen in, e.g., Ref.~\cite{ewan1997}, and it relies on the optically driven fluctuations in the matter fields. Also, for significantly larger condensates, with $N\approx10^6$, apparently chaotic classical dynamics are observed. It may be that this behavior persists even with the minimal ($N^{th}=0$) kicks given by the noise operators $\hat \xi$. To explore these possibilities, further work will be needed in the form of a full nonlinear quantum treatment.

\begin{acknowledgments}
This work is supported by the US National Science Foundation, the DARPA ORCHID program through a grant from AFOSR, and the US Army Research Office.

\end{acknowledgments}


\begin{thebibliography}{10}
\bibitem{kippenberg2008}
T. J. Kippenberg and K. J. Vahala, Science {\bf 321}, 1172 (2008).

\bibitem{marquardt2009}
F. Marquardt and S. M. Girvin, Physics {\bf 2}, 40 (2009).

\bibitem{oconnell2010}
A. D. O'Connell, et al., Nature {\bf 464}, 697 (2010).

\bibitem{teufel2011}
J. D. Teufel, et al.,  arXiv:1103.2144n1 (2011).

\bibitem{murch2008}
K. W. Murch, K. L. Moore, S. Gupta, and D. M. Stamper-Kurn, Nat. Phys. {\bf 4}, 561 (2008).

\bibitem{brennecke2008}
F. Brennecke, S. Ritter, T. Donner, and T. Esslinger, Science {\bf 322}, 235 (2008).

\bibitem{esslinger2009}
S. Ritter, F. Brennecke, K. Baumann, T. Donner, C. Guerlin, and T. Esslinger, Appl. Phys. B {\bf 95}, 213 (2009).

\bibitem{rina2010}
R. Kanamoto and P. Meystre, Phys. Rev. Let. {\bf 104}, 063601 (2010).

\bibitem{nagy2009}
D. Nagy, P. Domokos, A. Vukics, and H. Ritsch, Euro. Phys. J. D {\bf 55}, 695 (2009).

\bibitem{liu2009}
J. M. Zhang, F. C. Cui, D. L. Zhou, and W. M. Liu, Phys. Rev. A {\bf 79}, 033401 (2009).

\bibitem{keye2009}
W. Chen, K. Zhang, D. S. Goldbaum, M. Bhattacharya, and P. Meystre, Phys. Rev. A {\bf 80}, 011801 (2009).

\bibitem{dan2009}
D. S. Goldbaum, K. Zhang, and P. Meystre, arXiv:0911.3234 (2009).

\bibitem{aranya2009}
A. B. Bhattacherjee, Phys. Rev. A {\bf 80}, 043607 (2009).

\bibitem{Inouye1999}
S. Inouye {\it et al.}, Science {\bf 285}, 571 (1999).

\bibitem{Schneble2003}
D. Schneble {\it et al.}, Science {\bf 300}, 475 (2003).

\bibitem{Schneble2004}
D. Schneble {\it et al.}, Phys. Rev. A {\bf 69}, 041601(R) (2004).

\bibitem{Yoshikawa2004}
Y. Yoshikawa {\it et al.}, Phys. Rev. A {\bf 69}, 041603(R) (2004).

\bibitem{Moore1999}
M. G. Moore and P. Meystre, Phys. Rev. Lett. {\bf 83}, 5202 (1999); M. G. Moore and P. Meystre, Phys. Rev. A {\bf 58}, 3248 (1998).

\bibitem{Piovella2001}
N. Piovella, M. Gatelli and R. Bonifacio, Opt. Commun. {\bf 194}, 167 (2001).

\bibitem{Robb2005}
G. R. M. Robb, N. Piovella and R. Bonifacio, J. Opt. B {\bf 7}, 93 (2005).

\bibitem{Zobay2005}
O. Zobay and G. M. Nikolopoulos, Phys. Rev. A {\bf 72}, 041604(R) (2005).

\bibitem{Uys2007}
H. Uys and P. Meystre, Phys. Rev. A {\bf 75}, 033805 (2007).

\bibitem{Bonifacio1994}
R. Bonifacio and L. De Salvo, Nucl. Instr. Meth. Phys. Res. Sec. A {\bf 341}, 360 (1994).

\bibitem{Bonifacio21994}
R. Bonifacio {\it et al.}, Phys. Rev. A {\bf 50}, 1716 (1994).

\bibitem{Bonifacio1995}
R. Bonifacio and L. De Salvo, Opt. Commun. {\bf 115}, 505 (1995).

\bibitem{Bonifacio1997}
R. Bonifacio {\it et al.} Phys. Rev. A {\bf 56}, 912 (1997).

\bibitem{Lippi1996}
G. L. Lippi {\it et al.}, Phys. Rev. Lett. {\bf 76}, 2452 (1996).

\bibitem{Hemmer1996}
P. R. Hemmer {\it et al.}, Phys. Rev. lett. {\bf 77}, 1468 (1996).

\bibitem{Kruse2003}
D. Kruse {\it et al.}, Phys. Rev. Lett. {\bf 91}, 183601 (2003).

\bibitem{Cube2004}
C. von Cube {\it et al.}, Phys. Rev. lett. {\bf 93}, 083601 (2004).

\bibitem{chen2010}
W. Chen, D. S. Goldbaum, M. Bhattacharya, and P. Meystre, Phys. Rev. A {\bf 81}, 053833 (2010).

\bibitem{shore1990}
B. W. Shore, "The Theory of Coherent Atomic Excitation". Wiley, New York (1990), p. 824.

\bibitem{walls2007}
D. F. Walls and G. J. Milburn, "Quantum Optics, 2nd Ed." Springer, Berlin (2007).

\bibitem{tabor1989}
M. Tabor, "Chaos and Integrability in Nonlinear Dynamics: An Introduction". Wiley, New York (1989), p. 197.

\bibitem{Bhattacharya2008}
M. Bhattacharya and P. Meystre, Phys. Rev. A {\bf 78}, 041801(R) (2008).

\bibitem{ewan1997}
G. J. Milburn, J. Corney, E. M. Wright, and D. F. Walls, Phys Rev. A {\bf 55}, 4318 (1997).

\end{thebibliography}
\end{document}